\magnification=1200     
\font\small cmr10 scaled \magstep0

\def\newline{\hfill\break}
\def\blankline{\par\vskip 12 pt\noindent}
\def\mincir{\ \raise -2.truept\hbox{\rlap{\hbox{$\sim$}}\raise5.truept  
\hbox{$<$}\ }}                                                          %
\def\magcir{\ \raise -2.truept\hbox{\rlap{\hbox{$\sim$}}\raise5.truept  %
\hbox{$>$}\ }}                                                          %
\def\ref{\par\noindent\hangindent 20pt}
\hsize 16truecm
\vsize 22truecm

\vskip 6cm
\blankline
\blankline
\centerline{{\bf FAINT BLUE GALAXIES AND MERGING:}} 
\centerline{{\bf THE EVOLUTION OF THE LUMINOSITY FUNCTION }} 
\vskip 1cm

\centerline{{\it A. CAVALIERE $^1$, and N. MENCI $^2$}}
\smallskip
\centerline{$^1~$Astrofisica, Dipartimento di Fisica II Universit\`a di Roma,}
\centerline{via della Ricerca Scientifica 1, I-00133 Roma, Italy.}
\smallskip
\centerline{$^2~$Osservatorio Astronomico di Roma,}
\centerline{via dell'Osservatorio, I-00040 Monteporzio (Roma), Italy.}
\vskip 2cm
{\centerline{\it  ABSTRACT.} }
\vskip 0.4cm
We probe to what extent not only the counts of the 
faint blue galaxies and their redshift distribution but also their 
$z$-resolved 
luminosity functions can be explained in terms of binary merging 
(aggregations). 

We present a dynamical theory of such interactions. On this basis we 
find that ``minimal aggregations'' taking place 
within large scale structures and triggering starbursts
yield rates and timing such as to explain the observations of the local, flat 
 luminosity function  and of its progressive rising and 
steepening for redshifts out to 
$z\sim 1$. 
 Correspondingly, we predict faint blue counts still 
rising up to $m_B\sim 28 -29$, redshift distributions shifting toward larger 
 $z$ with increasing $m_B$; in addition, we predict 
an upturn of the faint end of the 
luminosity function  more pronounced in clusters than in the field. 

We propose that our picture provides the differential 
dynamics missing in the canonical 
hierarchical clustering theories. This reconciles with 
the observations 
the steep luminosity 
functions predicted at high $z$ by such theories. 

\vskip 1cm
{\it Subject headings:} cosmology --  galaxies: evolution -- galaxies:
formation.

\vfill\eject

1. {\it INTRODUCTION}
\blankline 
The observations of faint blue galaxies (FBGs) 
provided statistically 
significant counts in works by Koo (1986), 
Tyson (1988), and 
Cowie et al. (1988). Then an excess became apparent, 
namely, by a factor $\approx  5$
at magnitudes $B \mincir 24$ and $\approx 7$ at fainter ones, 
over the values expected from 
normal local galaxies filling a critical 
Friedmann-Robertson-Walker universe.
Over the last decade the 
growing data base concerning FBGs 
kept challenging the interpretations.

Various contributions to that excess have been discussed, the simplest being 
based on improved measurements of the local luminosity function 
(hereafter local LF). 
 These include the correction of the 
incompleteness by a factor $\mincir 1/2$ in the previous 
samples of local galaxies, as 
pointed out by Lilly et al. (1995) and by Ellis et al. (1996), 
and the consideration of the upturn at $M_B \geq -16$ from a flatter 
Schechter shape pinpointed by Marzke, Huchra, \& Geller (1994).
Both these improvements in the local baseline 
increase, albeit moderately, the expected counts, since the 
latter are given by the  convolution along the line of sight of the 
LFs weighted with the 
cosmological volumes. 

Meanwhile, it became apparent that toward  faint magnitudes 
blue galaxies make up a   
fraction of all galaxies which increases both with decreasing luminosity 
and with increasing redshift $z$ (Koo \& Kron 1992; Colless et al. 1993).
Thus, some sort of galaxy evolution must be at work. 

Passive luminosity evolution driven by the shift with age of 
stellar populations 
would provide the observed blue counts only if 
helped by the larger volume contributed by widely 
open  cosmologies with $\Omega_o\ll 1 $ (see Pozzetti, 
Bruzual, \& Zamorani 1996). 
A similar result also holds in the red band down to $K= 24$ 
(Cowie et al. 1994; Djorgovski et al. 1996), though the excess is 
considerably smaller here. Such cosmologies, however, are out of tune with other 
evidence (summarized, e.g., by Ostriker \& Steinhardt 1995) favoring  
 a flat universe of higher density, i.e., $\Omega_o\magcir 0.3$ with a 
cosmological constant $\lambda_o=1-\Omega_o$. 

On the other hand, a constraint was added with the advent 
of large spectroscopic samples 
(Cowie, Songaila, \& Hu 1991; Lilly 1993; 
Colless et al. 1993; Glazebrook et al. 1995) that provided redshift 
distributions $N(z)$ with a median value as low as 
$\langle z \rangle \approx 0.5$-0.6  
for the identified objects with  $B \mincir 24$.  
The low value of the redshift where $N(z)$ is peaked 
rules out strong luminosity evolution in a flat universe; rather, 
for $z \mincir 0.5$, the distribution resembles the behavior 
provided by passive-evolution models
with sophisticated color mixes (Koo 1990). The latter, on the other hand,  
would imply a very substantial high-$z$ tail unless 
very specific recipes are implemented (see Gronwall \& Koo 1995), which imply 
 mismatches at lower $z$ (see Cowie et al. 1996). 

However, recently Lilly et al. (1995) and Ellis et al. (1996)
 have provided 
{\it direct} evidence of considerable 
{\it evolution} in the resolved luminosity functions over the 
range  $z \approx 0.1 - 1$.  This concerns only the {\it blue} objects, 
 while the red ones appear 
to be already ``in place'' by $z\approx 1$, 
and quite similar to typical $L_*$ 
galaxies of the local kind mainly undergoing passive change. 
The observed evolution appears to be strongly 
differential, being faster for the lower luminosities measured. 

Two main interpretations have been proposed  for the onset of such a 
blue, later lineage 
of galaxies. One (Babul \& Rees 1992; Babul \& Ferguson 1996) 
holds these to be very small and numerous lumps in which 
gas collapse and star formation 
have been delayed by the high level of the UV background 
  prevailing before $z \approx 1$; shortly 
 thereafter, the photoionizing UV flux 
might decline sharply enough for this constraint to be lifted. 
Following the initial short starburst, these objects would rapidly fade away.

The other interpretation stems from the widespread local evidence 
of starbursts associated with galaxy-galaxy interactions 
(among the vast literature see the recent papers by Zabludoff et al. 1996; 
 Hutchings 1996 and references therein); this  
maintains that many more, if smaller, blue starbursts are 
triggered  at increasing depths by tidal interactions 
 taking place mainly with or between numerous dwarfs 
(Lacey  \& Silk 1991). 
Some such interactions are likely to end up in galaxy-galaxy aggregations, 
i.e., {\it binary merging}, also causing considerable number evolution.  
A phenomenological treatment of 
 these effects has been outlined by 
 Guiderdoni \& Rocca-Volmerange (1991)  and by Broadhurst, Ellis, \& Glazebrook 
(1992). 

A dynamical approach to the kinetics of aggregations 
has been given by Cavaliere \& Menci (1993), 
motivated to complement the standard hierarchical clustering theory 
by its intrinsic inadequacy in two specific respects: 
it predicts {\it too 
many} small objects, and it envisages {\it too little} 
dynamics to erase them; so 
by itself it would set and 
{\it preserve} a steep luminosity function. A reconciliation 
between {\it excess} counts and a 
locally {\it flat} LF may be sought in terms of 
 star formation processes (see White 1994; Frenk, Baugh, \& Cole 1995), 
but it still falls short of the full goal even on tuning several 
parameters. 
Cavaliere \& Menci (1993) instead, on the basis of 
the {\it dynamics} of aggregations, 
predicted that $N(M,z)$ 
 and hence $N(L,z)$ should flatten out with $z$ decreasing 
in the range $z\mincir 1$. 

We are encouraged to pursue our dynamical approach further not only by the 
recent  data concerning the $z$-resolved luminosity functions 
(which we now compare with) but also by  some 
 very recent lines of additional evidence. These include the following: 
a sizable fraction of FBGs  at $z\sim 1$ 
show dynamical signs of ongoing or recent interactions, 
or the presence of blue 
satellites (Koo et al. 1996; Van den Bergh et al. 1996); 
 some FBGs look like spirals with a red bulge (Koo et al. 1996) 
and peripheral starbursts; widespread post-starburst  galaxies 
mark the action of 
pervasive galaxy-galaxy interactions (Zabludoff et al. 1996); 
measurements of clustered  redshifts have shown the 
presence of at least some galaxy concentrations 
(Le F\`evre et al. 1994; Koo et al. 1996). 

Here we will start from first principles and build up a dynamical theory of 
interactions containing  no free parameters  once  
one specifies  the environment where 
the interactions take place. 
 Our main point will be that many such 
interactions can occur even at moderate redshifts ($z\sim 1$)  
 within large scale structures that offer the 
best combination of volume and density contrast. Such interactions, 
with their mass-dependent cross 
section $\Sigma(M)$, naturally yield a {\it differential} 
evolution. This affects the dwarf 
galaxy numbers yielding considerable {\it density} evolution, 
while the associated starbursts affect all blue luminosities 
yielding a milder {\it luminosity}  
evolution. In terms of observables, we find counts still rising at very 
faint magnitudes, redshift 
distributions peaked at a value shifting from $z\approx 0.5$ 
at $B\mincir 24$ toward 
$z\approx 1$ for fainter magnitudes, and resolved luminosity distributions 
which  that flatten out from $z \approx 2$ 
toward a local Schechter-like shape. 
 
The plan of the paper is as follows. In \S 2 we present our dynamical 
framework to 
 compute the mass functions evolving by aggregations.  
In \S 3 we discuss the 
 strong dependence of such evolution on the galaxy environment. 
 We compare three kinds of environments: the homogeneous ``field'';
 the bound, virialized structures; 
and the large-scale structures, which in fact will dominate the observables. 
 In \S 4 we describe how 
from the evolutionary mass distribution we compute the observables: 
the resolved luminosity distribution $N(L,z)$, 
the blue counts $N(m_B)$ and the redshift distribution $N(z)$. 
In \S 5 we give our results and predictions for these observables. 
 In the final section we give  and discuss our conclusions.

\blankline
{\it 2. THE MASS FUNCTION FROM BINARY AGGREGATIONS}
\blankline
{\centerline{\it 2.1 The dynamics}}
\blankline
A dynamical theory for galaxy evolution directly concerns the 
 mass function $N(M,t)$. The governing equation will be of the general form 
$${\partial N\over \partial t}\sim {N\over \tau}~,\eqno(2.1)$$
where $\tau^{-1}$ is an average evolutionary rate. 

In the simplest case when the evolution of a galaxy  
is not affected by surrounding companions, 
the rate will not depend on the mass function itself. 
But this 
produces self-similar $N(M,t)$ with faint-end slope 
 unchanged in time. 
A well known instance is constituted by the Press \& Schechter (1974) 
formula, corresponding to the choice of $\tau$ discussed by 
Cavaliere, Colafrancesco, \& Scaramella (1991) and by Kitayama \& Suto (1996). 

Hence,  we consider the case when 
the evolutionary rate $\tau^{-1}$ does depend on the number of 
surrounding masses that interact and eventually aggregate. 
For a total density $n_g(t)$ of galaxies, 
this corresponds to considering 
$\tau^{-1}\propto n_g\,\langle\Sigma\,V\rangle$,  
where $\Sigma$ is the cross section for aggregation and $V$ is the relative 
velocity of the two aggregating masses; the  average runs over the galaxy 
velocity distribution.  
The requirement of total {\it mass conservation} (see Lucchin 1988) 
requires a positive and a negative term on the right hand side of equation 
 (2.1), 
and easily leads to the following Smoluchowski form for the  kinetic equation: 
$$ {\partial N \over \partial t} = {1 \over 2} \int_0^m dm'~
K (m',m-m',t) ~N(m',t) ~ N(m-m',t) $$
$$ - N(m,t) \int_0^{\infty} dm'~
K(m,m',t) N(m',t) ~. \eqno(2.2)$$
Here the kernel 
$K(m,m',t)= n_g(t)\langle\Sigma\,V\rangle$ corresponds to 
the interaction rate for each pair of masses $m$, $m'$, and contains 
 the physics driving the evolution of the distribution $N(m,t)$. 
 The latter is written in its  comoving form;  the 
masses $m\equiv M/M_*$  are 
normalized in terms of $M_*$, the mass corresponding locally to the standard 
 characteristic luminosity $L_*$ (see Peebles 1993). 

Each galaxy aggregation comprises two stages: first, the dark matter halos 
 coalesce; second, the baryonic cores may eventually coalesce. 
The former process requires 
grazing, weakly hyperbolic 
encounters for which the cross section is given by (Saslaw 1986, 
Cavaliere, Colafrancesco, \& Menci 1992, hereafter CCM92) 
$$\Sigma=\epsilon({V\over v_m})\,\pi\,\big(r_m^2+r_{m'}^2\big)\,
\Big[1+{G\,(m+m')\over r_m\,V^2}\Big]~~~~~~
{\rm with}~~~~r_m=r_*\,m^{1/3}~,~~~v_m=v_*\,m^{1/3}~,\eqno(2.3)$$
where $v_*$ and $r_*$ are the velocity dispersion and dark-halo 
radius of the present
$M_*$ galaxy. We use $M_*=5\,10^{11}\,h^{-1}\,M_{\odot}$ {
\footnote{$^1$}{\small{
We parameterize the Hubble constant using $H_o=100\,h$ km/s/Mpc. 
}
}; 
$r_*$ is conservatively 
taken to be $40\,h^{-1}$ kpc (see Persic, Salucci, \& Stel 1996), and 
$v_*=270$ km/s (three dimensional). 
The function  $\epsilon (V/v_m)$ describes the decreasing efficiency of the 
aggregations for 
increasing relative velocities. Following the results from N-body simulations 
(see Richstone \& Malumuth 1983)
 we will take $\epsilon=1$ for $V<3\,v_m$, and just $\epsilon=0$ for larger values;  
we shall discuss the effect of such a cutoff in \S 5 and \S 6. 
The second term (proportional to $G$) in the cross section  describes the 
focusing effect of gravity, an important addition to the purely 
geometrical cross section proportional to 
$\pi\,r_m^2$ only for large masses and in the 
range $V<\sqrt{3}\,v_m$ that prevails in long-lived poor groups 
(see CCM92).  

The second stage is constituted by the possible 
coalescence of the baryonic cores. 
Sufficient conditions for this to occur have  been discussed extensively in 
terms of dynamical friction of one core against the background provided by 
the merged halos (see, e.g., Kauffmann, White, \& Guiderdoni 1993). The 
associated scale $t_f$ 
is at most a factor of 5 longer than the halo crossing time when 
halo aggregations are effective, namely, for grazing encounters 
with $V<3\,v_m$, as discussed above. 
In all such cases, we find that $t_f$ does not exceed the 
halo aggregation time $\tau$ 
which drives the evolution of $N(M,t)$ after 
 equation (2.1). Note that such a condition is a sufficient and  
conservative one, as discussed by Kauffmann et al. (1993), and 
as confirmed by detailed simulations (including hydrodynamics) 
of single collisional events and of the ensuing 
aggregation process (see Barnes \& Hernquist 1991, Barnes 1992).
 
The solutions of equation  (2.2) depend on the 
setting of the initial conditions, not only as a starting point of the 
integration procedure, but also -- in keeping with the quadratic nature of 
the right-hand side -- for establishing the 
amplitude 
of the kernel. We 
discuss these points in turn. 
\blankline
\centerline{\it 2.2 Initial conditions} 
\blankline
The initial conditions are set up
at the time $t_{in}$ corresponding to a redshift $z_{in}$. 
Because we think of aggregations as {\it complementary} to the standard 
hierarchical clustering (for which see Peebles 1993), we focus on initial 
 mass functions having the  Press \& Schechter (1974) form
$$N(m)~dm={2\,a\,b\,\delta_c\over \sqrt{\pi}M_*}\,
\phi(m)\,dm~,\eqno(2.4)$$
where $\phi(m)=m^{-2+a}\,exp({-b^2\,\delta_c^2m^{2a}/2})$. 
Here $\delta_c\approx 1.68$, $b^{-1}\sim 1$ fixes the amplitude,  
and $a=(n+3)/6$ is related to the 
spectral index  $n\sim -2$ of the power spectrum for the density fluctuations 
from which galaxies  form (see White 1994). 

In the same spirit, the typical galactic mass 
at the initial time $M_*(t_{in})$  is identified with the 
characteristic mass $M_c(t_{in})$ 
provided by the standard hierarchical theory; when normalized
 to its present value 
[e.g., $M_c(t_o)=0.6\times 
10^{15}\,h^{-1}\,M_{\odot}$ for $\Omega=1$], this 
reads (Lupini 1995) 
$${M_c(t_{in})\over M_c(t_o)} = 
\Bigg[{1+3{\lambda_o\over \Omega_o}\,\alpha^{-2} \over 
{t_o^2\over t_{in}^2} + 3{\lambda_o\over \Omega_o}\alpha^{-2} } \Bigg]^{1/3a}
\eqno(2.5) $$
for a density parameter $\Omega_o$ and for a 
cosmological constant satisfying the canonical relation 
$\lambda_o=1-\Omega_o$ characteristic of most inflationary scenarios (see 
Ostriker \& Steinhardt 1995), 
with $\alpha\equiv \pi\,H_o\,t_o/2\Omega_o^{1/2}$. 

In the following, we will consider mainly $b=1$, and $n=const\approx -2.5$ 
 on galactic scales. 

The initial mass function (2.4) not only provides the starting 
condition for equation  (2.2), but also affects 
the numerical value of the kernel 
at the initial time $t_{in}$. In fact, we obtain 
$$n_g(t_{in})= C\,\Omega_o\,\rho_{co} 
{2ab\delta_c\over 
\sqrt{2\pi}}\; {(1+z_{in})^3\over M_*(t_{in}) } ~ \int\,\phi(m)\,dm ~ , 
 \eqno(2.6)$$
integrated from $M_*/500$ (the smallest mass considered);  
here $\rho_{co}$ is the critical cosmological density at the present epoch, and 
$C$ makes explicit the contrast factor at $t_{in}$ of the environment where 
interactions take place.  

Note that assuming the Press \& Schechter expression (2.4) and the above  
values for the parameters is not mandatory;  
 because of 
the very negative values of the spectral index $n$ at galactic scales, 
the exponent of the power-law in $\phi(m)$ is 
always close to  -2, a generic 
feature of the mass distributions also found in other variants of the hierarchical 
clustering scenario 
(see, e.g., White 1994). On the other hand, shortly after $t_{in}$ the 
integro-differential nature of equation  (2.2) causes the solutions to lose memory of 
the detailed shape of the initial conditions. 
\vskip 1cm
\centerline{\it 2.3 Interaction rates} 
\blankline 
The numerical value of the interaction kernel 
$K=n_g(t)\langle\Sigma\,V\rangle$ 
determines the timescale for the aggregation process. 
Computing $n_g$ as described above, and substituting 
fiducial values for $r_*$ and for the initial relative velocity $V_{in}$, 
 we obtain an expression for $K$ that we conveniently write in the form    
$$K(m,m',t)=3\,10^{-4}\,H_o\,t_o\,\,
(m^{2/3}+m'^{2/3})\,
\Big[1+{v_*^2\over V^2}\,(m^{2/3}+m'^{2/3})\Big]\times $$
$$
\bigg({r_*\over 10~{\rm kpc}}\bigg)^2\bigg({V_{in}\over 100~{\rm km/s}}\bigg)
n_g(C,\Omega,t_{in})\,{f(\Omega_o,\lambda_o,t)\, / t_o} ~. 
\eqno(2.7)$$
The function $f(\Omega_o,\lambda_o,t)\equiv n_g(t)\,V(t)/n_g(t_{in})
\,V(t_{in})$ 
describes the change of the kernel after the initial $t_{in}$ (depending, we 
recall, on the choice of $\Omega_o$, $\lambda_o$), 
and includes the evolution 
of the relative velocities and that of the density 
$\rho(t)$ within the environment. 
The factor $t_o^{-1}$ makes explicit the unit of time we use 
throughout the computation. 

The initial density $n_g(t_{in})$ is given by equation  
(2.6). Its value depends on the cosmological scenario under consideration; 
e.g., for an initial time $t_{in}$ corresponding to a redshift $z_{in}=3$ 
and for a  CDM spectrum ($n\approx -2.2$ at galactic scales) 
with $\Omega=1$ and $C=1$,  we obtain 
 $M_*(t_{in})=5\times 10^{11}\,h^{-1}\,M_{\odot}$,   
so that $n_g(t_{in})\approx 1.8\,h^3\,(1+z_{in})^3\approx 1.2\times
 10^2~ h^3$ 
Mpc$^{-3}$ down to $M_*/500$. 

The solutions of equation  (2.2) are unique (McLeod 1962) 
once we have specified the
initial conditions (through eqs. [2.4] $-$ [2.6]) and the 
aggregation rate (equation  [2.7]). 
We stress that these in turn depend on the following quantities: 
 $C$, $t_{in}$, $V_{in}$, $f(\Omega_o,\lambda_o,t)$; their values describe the 
environments where aggregations are operating, as will be discussed next. 
\blankline
{\it 3. ENVIRONMENTS}
\blankline 
The kinetic equation (2.2) is naturally written 
in terms of $t$. In the following 
we will  derive in terms of $z$
the quantities entering this equation and the resulting observables; the 
conversion from $t$ to $z$ for flat universes may 
be found, e.g., in Peebles (1993). 

\blankline
\centerline{\it 3.1 The homogeneous ``field''}
\blankline
If any such description is realistic - which we doubt in view of the 
impressive evidence of large scale structures recalled below - 
the contrast factor should be taken as 
$C=1$; the density $n_g\propto (1+z)^{3}$ 
decreases after $t_{in}$ in keeping with the general expansion; 
the relative velocities, given an initial value, 
scale as $V\propto (1+z)$. 
Thus, the aggregation rate changes with $z$ according to 
$$f(z)=\bigg({1+z\over 1+z_{in}}\bigg)^4~,\eqno(3.3)$$
where $z_{in}$ is the redshift corresponding to the initial time $t_{in}$. 

The corresponding rapid decrease 
of $f(t)$ with time makes  the aggregation rate 
rapidly 
negligible, so that 
the evolution of the mass function after equation  (2.2) will become mild at most. 
 Nevertheless, for comparison purposes we compute it using a value for $z_{in}$ 
 typical of the formation era of galaxies, according to equation  (2.5) 
which in turn depends on $\Omega_o$, $\lambda_o$ and 
on the perturbation spectrum. For a scale-free spectrum with $n=-2.2$ 
 we obtain 
$z_{in}=3.5$ when $\Omega=1$ and $\lambda_o=0$; and $z_{in}=4.5$ when 
$\Omega_o=0.2$ 
and $\lambda_o=0.8$. The initial $V_{in}=V_o\,(1+z_{in})$ 
is given in terms of the local value 
$V_o$ for the pairwise galaxy velocity dispersion. 

The latter was measured first by Davis 
\& Peebles (1983) (see also Tormen et al. 1993)  who found $V_o\approx 
 340$ km/s. Recently Marzke et al. 1995 combined the data from the CfA 
 Redshift Survey and the Southern Redshift Survey and obtained a larger value 
 $\approx 540$ km/s; however, when the Abell clusters with $R\geq 1$ are 
removed  from the sample, the pairwise ``velocity dispersion of the remaining 
galaxies drops to $V_o=298$ km/s". We adopt the value 300 km/s. 
 Actually, such a value for $V_o$ is conceivably 
due to the present perturbed gravitational potential; so our 
scaling back as $1+z$ in this subsection will optimistically 
maximize the effect, 
which will be shown to be inadequate anyway. 


\blankline
\centerline{\it 3.2 Virialized structures}
\blankline
Following the standard hierarchical clustering, 
virialized structures like groups or  clusters as environments 
 for galaxy interactions may be characterized simply in terms of the 
redshift $z$ at virialization. Then, the  
 density contrast is $\sim 180$, the galaxy density scales as 
$n_g(z)/n_g(0)=(1+z)^3$, and the 
  virial velocity dispersion as 
$$V(z)/V_o=(1+z)^{n-1\over 2(n+3)}\eqno(3.4)$$  
for $\Omega=1$; here the spectral index takes on values 
in the range $n= -1.5$, $-1.3$ appropriate for 
the scale of groups and clusters. 
Thus, the average aggregation rate would scale as the simple product of these 
two $z$-dependencies. However, we have to consider the 
limited survival time of the structure before it is 
reshuffled into larger and less dense condensations. Such time 
is evaluated to be $\tau_p=3\,t/2$ on average 
(Cavaliere, Colafrancesco, \& Scaramella 1991; Lacey \& Cole 1993), which 
corresponds to $\tau_p=3\,t_{in}\,(1+z_{in})^{3/2}/2\,(1+z)^{3/2}$
 for $\Omega=1$ and $\lambda_o=0$. 
Including the suppression factor due to limited survival, 
the {\it effective} aggregation rate is $\tau^{-1}\,\tau_p/t_o$,  
and in full reads 
$$n_g\,\langle \Sigma\,V\rangle\,\tau_p/t_o=
n_g\,\langle \Sigma\,V\rangle(z_{in})\,
{3\over 2}\,{t_{in}\over t_o}\,
\Bigg({1+z\over 1+z_{in}}\Bigg)^{2n+1\over n+3}~.
\eqno(3.5)$$
This is equivalent to considering the scaling 
$$f(z)=\Bigg({1+z\over 1+z_{in}}\bigg)^{2n+1\over n+3}~,\eqno(3.6)$$
and an effective contrast 
$$C=270~t_{in}/\,t_o~.\eqno(3.7)$$

We start our computation at a 
redshift typical for formation of groups of galaxies, $z_{in}=2.2$
for an index $n=-1.3$ at the cluster scale. With this value, the initial 
$V_{in}$ is evaluated from equation  (3.4) on the basis of a 
(three-dimensional) value $V_o=1270$ km/s for clusters of richness 1, 
to obtain $V_{in}=480$ km/s. The 
corresponding effective contrast from equation  (3.7) is $C\approx 45$ for 
$\Omega =1$. 

\blankline
\centerline{\it 3.3 Large scale structures}
\blankline
The large scale structures (LSSs; among the vast literature see 
 De Lapparent, Geller \& Huchra 1992; Sahni \& Coles 1995; 
Vettolani et al. 1996)
constitute an interesting environment for interactions to take place in; 
over the field they have 
the advantage of a larger density, and over groups and clusters 
 the advantage of lower velocity dispersions. 
However, they are quantitatively less understood 
as yet from both the observational 
and the theoretical points of view. 

In  the following, we shall use the semi-analytical approach proposed 
by Do-\hfill\break
roshkevich et al. (1995) based on 
the Zel'dovich approximation and 
 supported by numerical N-body simulations (see again Doroshkevich et al. 
1995, and references therein). 
In such a picture, caustics form first in the matter field at $z_f\approx 4$ 
(for $\Omega=1$), then LSSs build up  
by accretion of matter falling onto them. 
 In this stage aggregations are not effective, because most 
galaxies have either closely parallel motions or large relative velocities.

Aggregation can be effective in a subsequent stage 
when the thickness of the pancakes approaches half of their 
typical separation. This is the epoch $z_{in}$ when we start our 
calculations. We evaluate it taking up the 
results by Doroshkevich et al. (1995), for $\Omega =1$.
For $z\mincir 3$ the 
mean comoving distance $\cal L$ between LSSs increases for decreasing $z$ 
after 
$${\cal L}_1\approx r_c\,\Bigg[{8\over 3 \pi^2}\,\sqrt{15\over 7}\,
(1+{y^2\over 2})\,exp(-{y^2\over 2})\Bigg]^{-1}~,\,\eqno(3.8)$$
where $y=(1+z)/(1+z_f)$ and $r_c$ depends on the power spectrum. 
The LSS thickness  in this stage turns out to be
$$\ell_1\approx\,2.2\,r_c\,\sqrt{1-{1+z\over 1+z_f}}~.\eqno(3.9)$$
According to the discussion above, we identify 
$z_{in}$ with the epoch when 
$\ell_1/{\cal L}_1$ is 1/2. 
This yields $z_{in}\approx 2.3$, independent of $r_c$.  

The final stage for $z<z_{in}$ constitutes a comparatively quiet era 
during which the behavior of the 
galaxy relative velocities flattens out (Menci \& Valdarnini 1993) 
to the observed value $V_{in}\approx 300$ km/s 
(see Davis \& Peebles 1983; Tormen et al. 1993). 
During this era, the scale of voids  evolves as ${\cal L}_2\approx 
\big[\sqrt{1+[(1+z_t)/(1+z)]^2} - 1\big]^{1/2}$ (Doroshkevich et al. 1995), 
with $z_t\approx 0.2$. From 
such an expression we find the comoving 
thickness of the structures to scale approximately as $(1+z)^{1/2}$ so that 
the contrast behaves like $C(z)\propto (1+z)^{(D-3)/2}$, where $D$ is the 
dimensionality  of the LSS  ($D=2$ for sheet-like structures, and $D=1$ for
filaments). In terms of the function $f(z)$ defined in equation  (2.7) this 
translates to 
$$f(z)=\Big[(1+z)/(1+z_{in})\Big]^{(3+D)/2}~.\eqno(3.10)$$

If the local contrast is $\approx 5$, consistent 
with observations (Geller \& Huchra 1989; Ramella, private 
communication) and with numerical simulations 
(Brainerd and Villumnsen 1991; Menci \& Valdarnini 1993), then from the scaling 
of $C(z)$ we infer an initial contrast $C = 5 \, (1+z_{in})^{(D-3)/2}$, 
with the 
value $C 
\approx 3$ for 
$D=2$. 

\blankline

In this section we have defined the quantities which enter equation  (2.2) for the 
computation of the evolving mass function in different environments. 
A summary of the resulting 
numerical values is given in Table 1. 
Next we describe how, from the mass 
function, we compute the observables like the 
luminosity function, the galaxy number 
counts and redshift distributions. 

\blankline
{\it 4. FROM $N(M,t)$ TO OBSERVABLES}
\blankline

Our basic assumption in deriving the observables is that interactions also 
trigger star formation (see, e.g., Zabludoff et al. 1996,  Hutchings 1996, 
and references therein).
Specifically, we conservatively assume that in the range of parameters 
 where aggregations are effective, they also induce and sustain 
star formation at a rate $s(z)$ equal to 
the rate of dynamical interactions $\tau^{-1}(z)$ acting 
on the available gas 
(see Lacey \& Silk 1991; Broadhurst, Ellis, \& Glazebrook 1992).

The associated luminosity for the 
dwarf galaxies is given by the 
relation $L\propto M_{gas}/\tau$, considering that $\tau$ 
exceeds the dynamical time of a single galaxy 
 and constitutes the limiting 
factor. 
Moreover, 
the deposition of  energy from Supernovae will be initially balanced by 
cooling,  but then for shallow 
potential wells it will go predominantly into gas expulsion. 
In this regime,  even more directly than in 
White \& Frenk (1991), the relation 
$M_{gas} \propto M\,v_m^2$ is obtained. 
 Since $\tau^{-1}$ has the same $M$ and $t$ (or $z$) dependencies as given 
 for $K$ by equation  (2.7), we obtain 
$$L/L_*=\big(M/M_*\big)^{\eta}~~~~~~~~{\rm and}~~~~~~~~
L_*(z)\propto f(z,\lambda_o,\Omega_o)~,\eqno(4.1)$$ 
where  $\eta=4/3$. Such a value  applies when the cross section (2.3) is purely 
geometrical and is consistent with the observational results by 
Kormendy (1990). 
 From equation  (4.1) the luminosity function $N(L)=N[M(L)]\,dM/dL$ is obtained.

Since our aim is to isolate the effects of {\it dynamical} evolution, we 
deliberately  simplify colors and $k$-corrections to the bones. So we 
identify the prompt luminosity $L$ with that 
in the blue band (see Broadhurst et al. 1992), and 
in the blue spectral region at moderate $z$ we assume a neutral $k$-correction 
(see Lilly, Cowie, \& Gardner 1991 and  Koo \& Kron 1992). We take an absolute 
characteristic magnitude $M_{B*}=-20$ (see Lilly et al. 1995; Ellis et 
al. 1996). 

The counts and the redshift distributions 
are found by convolving the comoving $N(L,z)$ 
with the cosmological volume ${\cal V}$, to obtain 
$$N(m_B)\,dm_B=\int^{z_{max}}_0\,dz\,{{d\cal V} 
\over dz}\,(1+z)^3\,N[L(m_b,z)]\,
{dL\over dm_B}\,dm_B~,\eqno(4.2)$$
where the relation between $L$ and $m_B$ is given, e.g., by 
Guiderdoni \& Rocca-Volmerange (1991),  and the cosmological volume for a flat 
universe with 
$\Omega_o+\lambda_o=1$ is given, e.g., by Peebles (1993). The range of 
integration over $z$ extends out to the maximum $z$ from which the 
magnitude $m_B$ can be observed. 

\blankline
{\it 5. RESULTS}
\blankline

In this section we present our results for the mass distribution, the 
$z$-depen-\hfill\break
dent luminosity function $N(L,z)$, the counts $N(m_B)$ and 
the redshift distributions $N(z)$ in different  magnitude ranges.  
These observables 
are derived from galaxy evolution driven by aggregations acting in 
the different environments described in \S 3. 

The plots we present are obtained through the following 
steps:\newline
i) For a given galaxy environment we compute the aggregation rate 
given by equation  (2.7); 
the parameters $C$, $t_{in}$ (or $z_{in}$), $V_{in}$ and the function 
$f(\Omega_o,\lambda_o,t)$ are discussed in \S 3, and summarized in Table 1.
\newline
ii) We numerically integrate equation  (2.2) from the initial time $t_{in}$ 
to the present time $t_o$. The time step is set at 
$\Delta t=(t-t_o)/500$ while the mass step is $M_*/500$. 
\newline
iii) Following the equations recalled in \S 4 we compute 
$N(L,z)$, $N(m_B)$ and $N(z)$. 
\newline
The results we obtain are compared with the observations concerning 
the faint blue counts, the $z$-distributions,  and the 
$z$-resolved LFs. 

\blankline
{\centerline{\it 5.1 The homogeneous ``field''}}
\blankline
For comparison, we put on record the corresponding results, 
although this 
is not expected to be an environment conducive to many late binary 
 aggregations due to the rapid decrease of density and velocities. 

Figure 1a shows that such is indeed the case for $\Omega=1$ ($\lambda_o=0$). 
Because of the fast decrease of $f(\Omega_o,\lambda_o,t)$ 
with time (see Table 1), the interaction rate rapidly becomes small, 
so that $N(M,t)$ hardly evolves from the initial shape. 
The corresponding LFs are shown in 
Figure 1b. Any small evolution that occurs takes place 
 at high redshifts, so that no appreciable 
change can be observed in the range 
$z\mincir 1$. The faint counts in Figure 1c 
are well fitted, which is not not surprising considering the steep LFs at 
 all redshifts. But the $z$-distributions (Figure 1d) 
are peaked at $z\approx 1$ even at relatively  bright magnitudes. 

Thus, aggregations of galaxies 
uniformly distributed in a FRW universe are uninteresting, and clearly  
fail to account for the observed evolution in the LFs 
 out to $z\approx 1$. 

One might expect that the plateau (actually the inflection) characteristic 
of the cosmic scale  factor in a Lema\^itre universe ($\lambda\neq 0$) could 
 provide an era of roughly constant density when the binary aggregation 
processes could proceed for a while. However, in a flat universe 
such processes are 
 suppressed after the inflection by the accelerated 
expansion. The net effect is still close to the case 
$\Omega=1$, as can be seen 
from Figures 2a $-$ 2d computed for $\Omega_o=0.2$ and $\lambda_o=0.8$.  

\blankline
{\centerline{\it 5.2 Virialized structures}}
\blankline
A more favorable environment for galaxy aggregations 
is constituted by the bound, virialized structures like groups and clusters 
of galaxies. Here the large density contrast $C$ (though convoluted with the  
limited survival time  discussed in \S 3.2),  and 
the milder time decrease of the 
interaction kernel $f(\Omega_o,\lambda_o,t)$ yield a sustained 
evolution of $N(M,t)$ (see Figure  3a) compared to that found for the ``field''. 
In Figures 3b $-$ 3d we show 
the LFs, the counts, and the $z$-distributions 
expected if all galaxies resided within virialized structures for a 
considerable fraction of their lifetime. 
It is found that the evolution of $N(L,z)$ consists of a rapid flattening 
in the sub-$L_*$ range,  
 with relatively little change in the range from 
 $z\approx 1$ to the present. 

It must be noted that in clusters, in spite of the larger value of the 
contrast $C$, the aggregation rate is  suppressed not only by 
the limited survival time (see \S 3.2), but also 
by the increase with decreasing $z$ of the velocity dispersion entering the 
cutoff $\epsilon (V/v_m)$  in equation  (2.3). The latter circumstance has the 
effect of decreasing the gravitational term in $\Sigma$, and even of 
 shrinking to zero the whole cross section of the 
smaller galaxies with the lower internal velocities; so 
 at the faint end the mass distribution will hardly be changed from the initial 
steep shape. The net result will be a LF with an upturn toward  
the faint end (see Figure  3e). 
The balance between the effect of the sharp,  
mass-dependent decrease of the cross section
and the general speeding up driven by the  large effective 
contrast is a delicate one, and deserves an aimed study motivated by the 
LFs recently found (see Driver et al. 1994) 
to be appreciably steeper in local clusters than in the ``field''. 
 
\blankline
{\centerline{\it 5.3 Large scale structures} 
\blankline
A result in a way intermediate between the two 
extremes above is provided by the 
environment constituted by the LSSs, where observations show 
most galaxies to reside not only locally 
(Geller \& Huchra 1989), but also at increasingly larger redshifts (see 
 Broadhurst et al. 1990; Schectman et al. 1995; Vettolani et al. 1995; 
Carlberg et al. 1996).
 For galaxies inside sheet-like structures (our case $D=2$) 
 the parameters entering the aggregation rate are summarized in Table 1, and 
yield the evolution of $N(M,t)$ shown in Figure  4a. 
The corresponding counts and  redshift distributions 
(Figures 4c, 4d) agree with the observations. In particular, 
 they  agree with the 
 spectroscopic redshift distribution by Glazebrook at al. (1995) if 
$L\propto M^{4/3}$ holds, and yield an enhanced high-$z$ tail 
 for a stronger dependence of $L$ on $M$, e.g., $L\propto M^{1.5}$. 
 The luminosity functions (Figure  4b) show a clear flattening 
 from $z\approx 1$ toward a Schechter-like, local shape with 
$\alpha\approx -1.1$  in agreement with the data. 

Note that an upturn at faint luminosities should also be expected here; the  
 mechanism is similar to that discussed at the end of \S 5.2, but the 
effect is considerably milder (and shifted toward fainter luminosities) 
because the 
relative velocities inside LSSs are and remain smaller than 
those in clusters. 

Defining a characteristic luminosity by the ratio  of the second 
to the first moment of 
$N(L,z)$, we find this evolves by about 1 from $z=1$ to $z=0$, 
as also illustrated in Figure  4a. 
Actually, the intrinsic luminosity change $L_*(z)\propto (1+z)^{(3+D)/2}$ 
(see eqs. [4.1] and [3.10]) is nearly balanced by the 
change following approximately $(1+z)^{-1.5}$, due 
to characteristic mass after the relation $L\propto M^{4/3}$. The net 
result is a moderate change (about 1 Mag) of all luminosities, competing  with  the 
large change in number for setting the shape 
of $N(L,z)$. The corresponding $M/L$ ratio 
evolves (for an $L_*$ galaxy) as $M_*(z)^{-1/3}\,(1+z)^{-(3+D)/2}$ which 
yields $M_*/L_*\sim (1+z)^{-(2+D)/2}$. 

\blankline
{\it 6. CONCLUSIONS AND DISCUSSION}
\blankline

We have derived for FBGs a luminosity function 
that {\it changes} considerably for increasing $z$ as a result of 
{\it number} evolution and a milder {\it luminosity} evolution. 
In particular, in the observed range out to $z\approx 1$ the luminosity 
function $N(L,z)$ {\it rises} and {\it steepens} from a 
local logarithmic slope about $-1.1$ to a value close to $-1.6$ at 
redshift $\approx 0.8$, as observed (see Figure  4b, and below for a 
discussion). 

This is derived from a {\it dynamical} theory 
of the galaxy mass  function based on mass-conserving binary aggregations. 
The resulting evolution of $N(M,t)$ is far from being self-similar; 
indeed, it is strongly {\it differential} with mass, 
causing for smaller galaxies stronger number changes 
and shift of the mass scale. This constitutes an important 
complement to the canonical 
 hierarchical clustering, which in fact contains too little dynamics 
to account for the observed evolution of the luminosity function, 
 as recalled in the Introduction. 
 
The actual strength of the evolution induced by aggregations 
depends on the environment where these  
take place. In CCM92 
we stressed that in long-lived groups strong, gravitationally 
focused  interactions of {\it large} galaxies occur,  
leading to the formation of cD-like mergers. Here, 
we find that in the large-scale structures ---  
which comprise most local galaxies, are being observed 
 out to $z\sim 1$,  
and in a critical universe 
are expected to loom out at redshifts $z\sim  2$ ---  
{\it small-small} and {\it small-large} aggregations are favored. Their 
strength is such as to 
match -- with no special effort at optimization -- 
the existing data for the resolved $N(L,z)$ and 
for the integrated but deeper observables like counts and magnitude 
distributions (see Figures $4a$ to $4d$). 

A simple way of summarizing the interaction 
(and starburst) rates derived in \S 3.3 is to note that they scale --  
differently than in the ``maximal merging'' model, see Carlberg (1996)  -- 
like $\tau^{-1} = \tau_o^{-1} ~ (M_*/M)^{1/3} ~
(1+z)^{2.5}$. This is because in the rate $\tau^{-1}=n_g \Sigma V$ the scaling 
 of the background density $(1+z)^3$ is partly offset by the LSS contrast 
$C(z)\propto (1+z)^{-0.5}$; with mass, on the other hand, the scaling results 
 from $n_g\propto M^{-1}$ and $\Sigma\propto M^{2/3}$. 
 Toomre (1977) has fixed the observed interaction 
probability $t_o/\tau_o$ at  a few percent for interactions of 
bright local galaxies; we find $\tau_o^{-1}=C\,n_g(t_o)\,\Sigma\,V_o\approx 
1/500 \,~{\rm Gyr}^{-1}$. 
At $z \approx 1$ our probability $t/\tau$ for galaxies of $ 5\times 
10^9\,M_{\odot}$ 
 exceeds 30 \%. The bright career of this dwarf population activated 
 by interactions is effectively started at $z\approx 2$ when the 
large scale structures begin to loom out; they reshuffle the conditions 
set by the canonical hierarchical clustering including the initial 
correlations of linear perturbations, 
sustain contrasts larger than unity and produce non-linear relative 
velocities confined to one or two dimensions. This career 
terminates, due to the Hubble expansion, with a 
progressive fading for $z \mincir 0.5$ and with the 
 number of sub-L* units reduced by $\sim 10^{-1}$. 

The LSS features used in our computation 
-- filamentary ($D=1$) or sheet-like ($D=2$ as considered above)  
structures produce similar results --
come down to three key points: formation era around 
$z_{in}\approx 2$, local density contrast $\approx 5$, and 
slow growth of the contrast from $z_{in}$ to the present.  
These features, 
which appear in the relatively local observations and also in 
all N-body experiments, 
 are also incorporated in analytic interpretations different from those of 
Zel'dovich's (see Bond, Kofman, \& Pogosyan 1996). 
We stress that all such structures are ultimately traced back to 
 long-distance correlations, 
not considered  in the homogeneous renditions 
of the hierarchical clustering scenario, 
 but  actually intrinsic, and 
effective in bringing to interact galaxies that otherwise 
would not have met. 

The detailed value of the initial contrast $C$ is the only ``free'' 
parameter at present. 
In fact, its magnitude is constrained by its very definition 
to be of the order of a few, as we have used. 
As to finer adjustments, its local distribution 
is currently being measured (Ramella, private communication); 
the analytic theory 
of its evolution is being refined, 
beyond the approximation used in \S 3.3 (see 
Sahni \& Coles 1995); observations are reaching beyond $z\approx 1$ 
(see Willinger et al. 1996). 

Our results are conserved when the shape of 
the initial mass function is changed from the detailed Press \& Schechter 
 form but is still steep as predicated by all hierarchical clustering scenarios
 (cf. White 1994). 

The main trends are also robust with respect to variations of the $M/L$ ratio 
 from the value $M/L\propto M^{-1/3}$ used in \S\S 4 and 5, to include 
 $M/L\propto$ const. and  $M/L\propto  M^{-2/3}$ as shown in Figure  5. 
 The different $M/L$ ratios tilt, but moderately, the shape 
 of the distant LFs; correspondingly, the tail of $N(z)$ for $z>1$ rises 
 to the values indicated by the data by Cowie et al. (1996) if the dependence 
of $M/L$ goes toward $M^{-2/3}$.   

We also computed the counts in the $K$-band. These, as expected, increase 
considerably less 
than in the blue due to the long evolutionary times of the red stars and to 
the stronger  $k$-correction. 

One recurrent 
objection to the aggregation scenario concerns the 
correlation function, as discussed by Efstathiou (1995). 
This is a delicate issue, as is shown in Figure  6. 
At the statistical level of the correlation function, 
aggregations with cross section less than linear in mass
 (like $\Sigma\propto M^{2/3}$ for nearly geometrical cross sections 
that apply here, see 
\S 2) do not cause the slope of the initial correlation function to change.
A steepening instead takes place for the stronger, 
 gravitationally 
focussed interactions with non-linear cross section $\Sigma\propto M^{4/3}$.  
The whole issue may peter out, 
however, with the growing evidence for clustered redshifts at 
$z\approx 0.8 - 1 $ 
as observed 
and discussed by Le F\`evre et al. (1994) and by Koo et al. (1996). 

The other recurrent objection to aggregations is that the spiral disks may be 
damaged by merging, as discussed by Ostriker (1990). 
In our framework, however, at the level of $M_*$ 
the mass growth is due to minor episodes of merging with the 
numerous but much smaller (and often more diffuse) galaxies. 
 On the other hand, 
the overall change of the characteristic 
mass seen in Figure  4a is mainly due to the changing shape of the 
mass function, i.e., to its flattening 
corresponding to the disappearance of small galaxies.  
This results in a mild {\it statistical} change in time of the 
 second moment of the distribution, which does not imply a corresponding 
change in the masses of individual galaxies. For similar reasons, the 
population of ellipticals  does not evolve appreciably by aggregations in the 
redshift range considered here. 

To conclude, excess counts at $B\mincir 24$ may result from the combination 
of several effects: local shape and normalization of the luminosity function, 
 long line-of-sight associated with $\Omega_o<1$, 
 passive evolution, and interactions as stressed 
here. 
The shape of $N(z)$ is more telling, and  at relatively bright magnitudes 
$m_B\mincir 24$ requires some evolution, yet it rules out the 
pure luminosity kind in a flat universe. 
The resolved $N(L,z)$ with its evolving shape restricts the issue to 
the following alternatives: 
luminosity-dependent luminosity evolution or a combination of 
(differential) density plus luminosity evolution, as naturally 
provided by aggregations. The steepening associated with the latter drives 
a specific differential shift of $N(z)$, 
characterized by the peak marching on toward higher $z$  
 at fainter magnitudes (see Figure  $4d$); it also sustains  
the rise of the counts  at very faint magnitudes 
(see Figure  4c) even beyond $b_j=27$ measured by Metcalfe et al. (1995), 
probing in an integrated fashion the driver's 
action at redshifts not   
yet reached by spectroscopy. 

The above predictions stem from  simple dynamics 
 in which interactions produce aggregations and starbursts 
 together. On the other hand, Cavaliere \& Menci (1993) and Moore et al. (1996) 
point out that in clusters higher densities but larger relative velocities 
cause more frequent but weaker encounters (``harassment'') 
resulting mainly in pure 
starbursting and in blueing of the Butcher-Oemler (1984) type. 
 In fact, such flybys
and the true aggregations constitute two facets of a single process, 
differing as for the detailed form of 
the cross section (see equation  2.3) and for the ratio $s(z)/\tau^{-1}(z)$ 
of the star formation to the interaction rates. 

These two facets stem from the same dynamics, and
 concur to stress the importance of interactions as {\it drivers}
of galaxy evolution. Actually, the intermediate regime 
should also be observable in clusters, in the form of a local LF frozen 
with a steeper 
faint end or a more pronounced upturn, as shown by Figure  3e.
\blankline
We have benefited from many informative discussions with 
G. De Zotti, M. D'Onofrio, E. Giallongo, M. Ramella, R. Windhorst 
 and G. Zamorani. We thank the referee 
for his helpful comments which improved our presentation. This work 
was supported 
by partial grants from MURST and ASI.

\vfill\eject

{\it REFERENCES}
\blankline

\ref Babul, A., \& Rees, M.J. 1992, MNRAS, 255, 346
\ref Babul, A., \& Ferguson, H.C. 1996, ApJ, 458, 100
\ref Barnes, J.E., \& Hernquist, L.E. 1991, ApJ, 370, L65
\ref Barnes, J.E. 1992, ApJ, 393, 484
\ref Bond, J.R., Kofman, L.,  \& Pogosyan, D. 1996, Nature, 380, 603
\ref Brainerd, T.G., \& Villumsen, J.V. 1992, ApJ, 394, 409  
\ref Broadhurst, T.J., Ellis, R.S., Koo, D.C., \& Szalay, A.S. 1990, Nature, 
311, 726 
\ref Broadhurst, T.J., Ellis, R.S., \& Glazebrook, K. 1992, Nature, 355, 55
\ref Butcher, H., \& Oemler, A. 1984, ApJ, 285, L45
\ref Carlberg, R.G. 1996, in ``Galaxies in the Young Universe'', MPI 
Conference, ed. H. Hippelein (Springer), preprint
\ref Carlberg, R.G., Cowie, L.L., Songaila, A., \& Hu, E.M. 1996, 
preprint [astro-ph/9605024] 
\ref Cavaliere, A., Colafrancesco, S., \& Scaramella, R. 1991, ApJ, 380, 15  
\ref Cavaliere, A., Colafrancesco, S., \& Menci, N. 1992, ApJ, 392, 41 (CCM92) 
\ref Cavaliere, A., \& Menci, N. 1993, ApJ, 407, L9
\ref Colless, M.M., Ellis, R.S., Broadhurst, T., Taylor, K., 
\& Hook, R.N. 1993, MNRAS, 261, 19
\ref Cowie, L.L., Lilly, S.J., Gardner, J.P., \& McLean, I.S.  1988, ApJ, 332, 
L29
\ref Cowie, L.L., Gardner, J.P., Hu, E.M., Songaila, A., Hodapp, K.W., \& 
Wainscoat, R.J. 1994, ApJ, 434, 114 
\ref Cowie, L.L., Songaila, A., \& Hu, E.M. 1991, Nature, 354, 460
\ref Cowie, L.L., Songaila, A., Hu, E.M., \& Cohen, J.G. 1996, AJ in press 
[astro-ph/9606079]
\ref Davis, M., \& Peebles, P.J.E. 1983, ApJ, 267, 465
\ref de Lapparent, V., Geller, M.J., \& Huchra, J.P. 1992, ApJ, 369, 273
\ref Djorgowski, S. et al. 1996, ApJ, 438, L13
\ref Doroshkevich, A.G., Fong, R., Gottl\"ober, S., 
M\"ucket, J.P., \& M\"uller, F.  1995, preprint   
\ref Driver, S.P., Phillips, S., Davies, J.I., Morgan, I., \& Disney, M.J. 
1994, MNRAS, 266, 155
\ref Efstathiou, G.P.E. 1995, MNRAS, 272, L25
\ref Ellis, R.S., Colless, M., Broadhurst, T., Heyl, J., \& Glazebrook, K.   
1996, MNRAS, 280, 235
\ref Frenk, C.S., Baugh, C.M., \& Cole, S. 1995, in Proc. of `Mapping, 
Measuring and Modelling the Universe', Valencia, to appear 
\ref Geller, M.J., \&  Huchra, J.P. 1989, Science, 246, 897 
\ref Glazebrook, K., Ellis, R.S., Colless, M.M, Broadhurst, 
T.J., Allington-Smith, J.R., Tamvir, N.R., \& Taylor, K. 
1995, MNRAS, 273, 157
\ref Gronwall, C. \& Koo, D.C., 1995, ApJ, 440, L1 
\ref Guiderdoni, B., \& Rocca-Volmerange, B. 1991, A\& A, 252, 435
\ref Kauffman, G., White, S.D.M, \& Guiderdoni, B., 1993, MNRAS, 264, 201
\ref Kitayama, T., \& Suto, Y. 1996, MNRAS, 280, 638
\ref Koo, D.C. 1986, ApJ, 311, 651
\ref Koo, D.C. 1990, in ASP Conf. Ser. 10, `Evolution of the 
Universe of Galaxies - The Edwin Hubble Centennial 
Symposium', Aed. R.G Kron (San Francisco: ASP),  268
\ref Koo, D.C., \& Kron, R.G. 1992, ARAA, 30, 613
\ref Koo, D.C. et al. 1996, ApJ, 469, 535 
\ref Kormendy, J. 1990, in ASP Conf. Ser. 10, 
`Evolution of the Universe of Galaxies', 
 ed. R.G. Kron (San Francisco: ASP), 33
\ref Jones, L.R., Fong, R., Shanks, T., Ellis, R.S., \& Peterson, B. 1991,
 MNRAS, 249, 481
\ref Hutchings, J.B. 1996, ApJ, 111, 712
\ref Lacey, C., \& Cole, S. 1993, MNRAS, 262, 627 
\ref Lacey, C, \& Silk, J. 1991, ApJ, 381, 14
\ref Le F\`evre, O., Crampton, D., Hammer, F., Lilly, S.J., 
\& Tresse, L. 1994, ApJ, 423, L89
\ref Lilly, S.J., Cowie, L.L., \& Gardner, J. 1991, ApJ, 369, 79
\ref Lilly, S.J. 1993, ApJ, 411, 501
\ref Lilly, S.J., Tresse, L., Hammer, F., Crampton, D., \& 
Le F\`evre, O. 1995, ApJ, 455, 108
\ref Lucchin, F. 1988, in  Lecture Notes in Physics, 
332, `Morphological
Cosmology', ed. P. Flin (Berlin: Spriger),  284  
\ref Lupini, F. 1996, II Rome University Thesis
\ref Maddox, S.J., Sutherland, W.J., Efstathiou, G., Loveday, J., \& 
Peterson, B.A. 1990, MNRAS, 247, 1p
\ref Marzke, R.O., Huchra, J.P., \& Geller, M.J. 1994, ApJ, 428, 43
\ref Marzke, R.O., Geller, M.J., da Costa, L.N., \& Huchra, J.P. 1995, AJ, 
110, 477
\ref McLeod, J.V. 1962, Quart. J. Math. Oxford, 13, 119   
\ref Menci, N., Colafrancesco, S., \& Biferale, L. 1993, Journ. de Physique, 
 3, 1105
\ref Menci, N., \& Valdarnini, R. 1993, ApJ, 436, 559
\ref Metcalfe, N., Shanks, T., Fong, R., \& Roche, N. 1995, MNRAS, 273, 257 
\ref Moore, B. Katz, N., Lake, G., Dressler A., \& 
Oemler, A. 1996, Nature, 379, 613
\ref Ostriker, J.P. 1990, in ASP Conf. Ser. 10, 
``The Evolution of the Universe of Galaxies'', 
ed. R. Kron (San Francisco: ASP), 25
\ref Ostriker, J.P., \& Steinhardt, P. 1995, Nature, 377, 600
\ref Peebles, P.J.E. 1993, 
{\it Principles of Physical Cosmology} (Princeton: Princeton Univ. Press)  
\ref Persic, M., Salucci, P., \& Stel, F. 1996, MNRAS, 281, 27
\ref Pozzetti, L., Bruzual, G., \& Zamorani, G. 1996, MNRAS, 281, 953 
\ref Press, W.H., \& Schechter, P. 1974, ApJ, 187, 425 (P\&S)
\ref Richstone, D.O., \& Malumuth, E.M. 1983, ApJ, 268, 30
\ref Sahni, V., \& Coles, P. 1995, Phys. Rep., 262, 1
\ref Saslaw, W.C. 1985, `Gravitational Physics of Stellar and Galactic Systems'
(Cambridge: Cambridge Univ. Press), 231
\ref Shectman, S.A., Landy, S.D., Oemler, A., Tucker, D.L., Kirshner, 
R.P., Lin, H., \& Schechter, P.L. 1995, 
in ``Wide Field Spectroscopy and the Distant Universe'', 
in Proc. of 35th Herstmonceaux Conf., eds. S.J. Maddox and 
A. Aragon-Salamanca   (Singapore: World Scientific), 99 
\ref Toomre, A. 1977, in `Evolution of Galaxies and
Stellar Populations', ed. B.M. Tinsley \& R.B. Larson 
(New Haven: Yale Univ. Observatory), 401
\ref Tormen, G., Moscardini, L., Lucchin, F. \& Matarrese, S. 1993,
ApJ, 411, 16 
\ref Tyson, A.J. 1988, AJ, 96, 1
\ref van den Bergh, S.,  Abraham, R.G., Ellis, R.S., Tanvir, N.R., 
Santiago, B.X., \& Glazebrook, K. 1996, AJ, 112, 359 
\ref Vettolani, G. et al. 1995, 
in ``Wide Field Spectroscopy and the 
Distant Universe'', in Proc. of 35th Herstmonceaux Conf., 
eds. S.J. Maddox and A. Aragon-Salamanca  
(Singapore: World Scientific), 115 
\ref White, S.D.M., \& Frenk, C.S., 1991, ApJ, 379, 52    
\ref White, S.D.M. 1994, preprint MPA 831
\ref Willinger, G.M., Hazard, C., Baldwin, J.A., \& 
McMahon, R.G. 1996, ApJS, in press
\ref Zabludoff, A.I., Zaritsky, D., Lin, H., Tucker, D., Hashimoto, Y., 
 Schectman, S.A., Oemler, A., \& Kirshner, R.P. 1996, ApJ, 466, 104

\vfill\eject

\noindent
{\it FIGURE CAPTIONS}
\blankline
\ref Figure  1: Results for galaxies with $M/L\propto M^{-1/3}$ 
 interacting in the homogeneous 
``field'' with $\Omega=1$ and $\lambda=0$. \newline
Panel a): The solid lines show the comoving mass function 
at 10  equally spaced time intervals, from the initial
 time $t_{in}$ (dashed line) to the present.  
 \newline
Panel b): The corresponding luminosity 
functions in the B band for three redshift bins: $z=0-0.2$, 
 $0.2-0.5$ and $z>0.5$; the data are taken from Ellis et al.  
1996. \newline
Panel c): The computed galaxy counts (solid line), compared
 with data from various authors: Maddox et al. (1990) 
solid circles; Jones et al. (1991), open triangles; 
Tyson (1988), open circles; Lilly, Cowie \& Gardner (1991), 
 open stars; Metcalfe et al. (1995), solid triangles.  
\newline
Panel d): The computed redshift distribution in the 
magnitude range $22.5<m_B<24$ 
is compared with the data from Glazebrook et al. (1995).
\blankline
\ref Figure  2: Same as Figure 1 but for $\Omega_o=0.2$ and 
$\lambda_o=0.8$. 
\blankline
\ref Figure  3: Same as Figure 1 but for galaxies within 
virialized structures. 
\blankline
\ref Figure  3e: The effect of varying the position of the cutoff 
for the efficiency $\epsilon(V/v_m)$ in the cross section 
(2.3). In this figure $\epsilon=0$ is taken for $V>2.5\,v_m$ 
rather than $V>3\,v_m$ as adopted throughout 
the paper. 
\blankline
\ref Figure  4: Same as Figure  1 but for galaxies within 
large scale  structures with $D=2$. In panel b) we show also our prediction 
for the luminosity function at $z=1$ (dot-dashed line). In panel c) we show 
also our predictions for $24<m_B<26$. 
\blankline
\ref Figure  5: The evolution of the luminosity function 
of interacting galaxies within large scale structures, 
computed for $M/L=$constant (left panel) and for 
$M/L\propto M^{-2/3}$ (right panel). 
\blankline
\ref Figure  6: The time evolution of the slope $\gamma$ of the 
two-point correlation function for aggregating galaxies, 
as computed from a typical Montecarlo simulation (see Menci,
 Colafrancesco, \& Biferale 1993 for details); $\gamma_{in}$ is the 
 initial value. 
The dotted line shows the result for a cross section scaling as 
$\Sigma\sim M^{4/3}$, and 
 the solid line that for $\Sigma\sim M^{2/3}$. 
\blankline

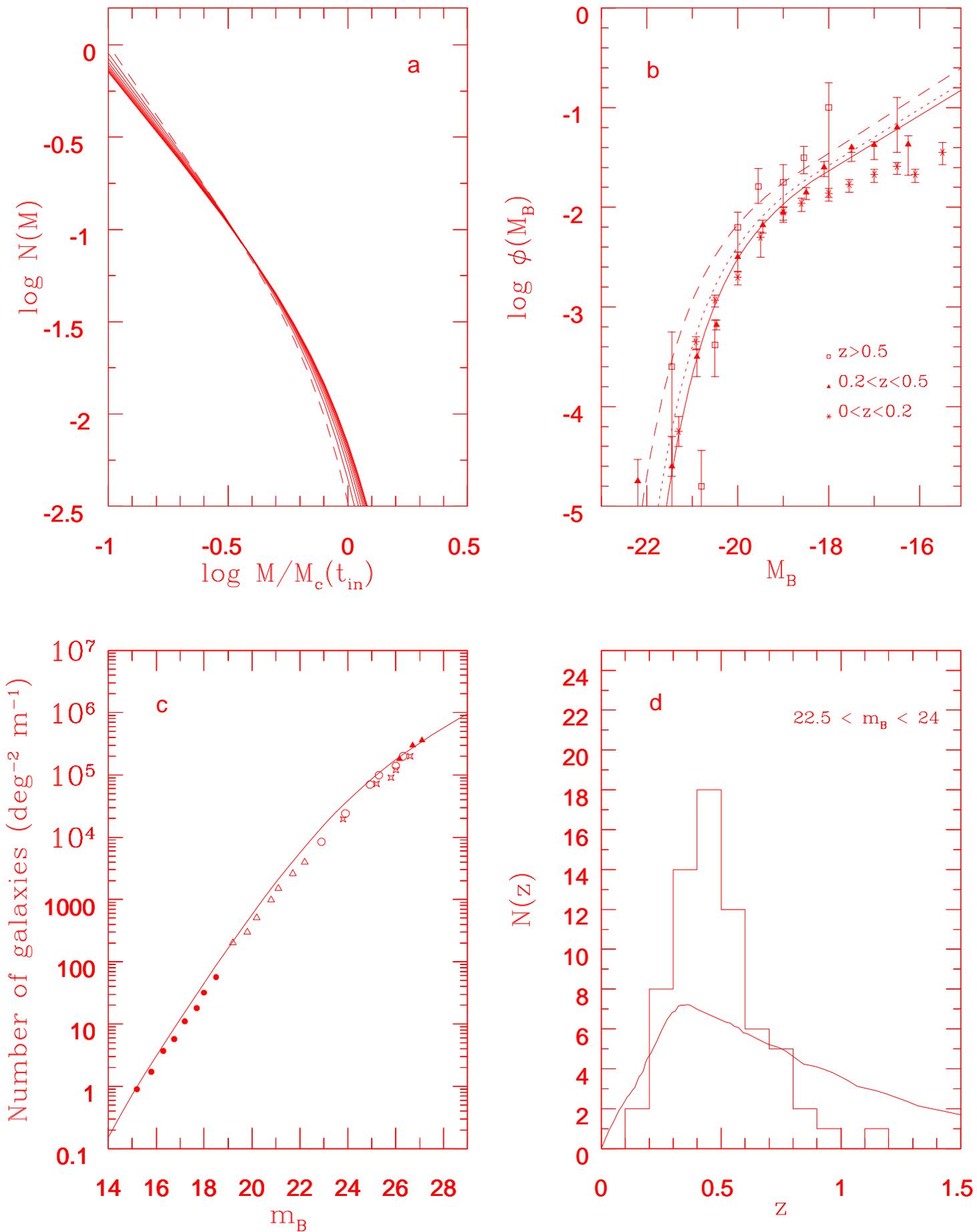
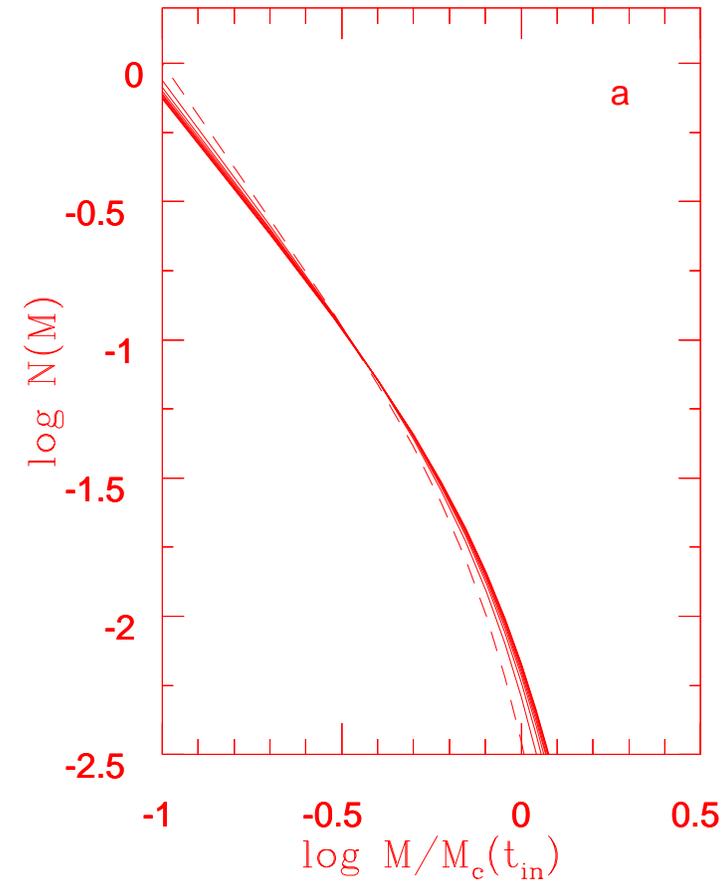
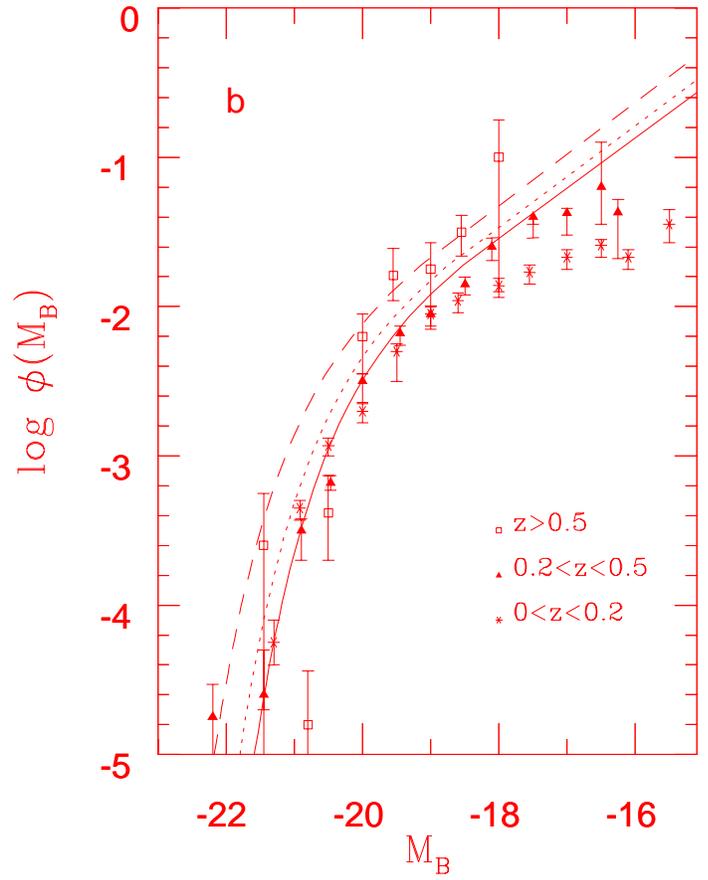
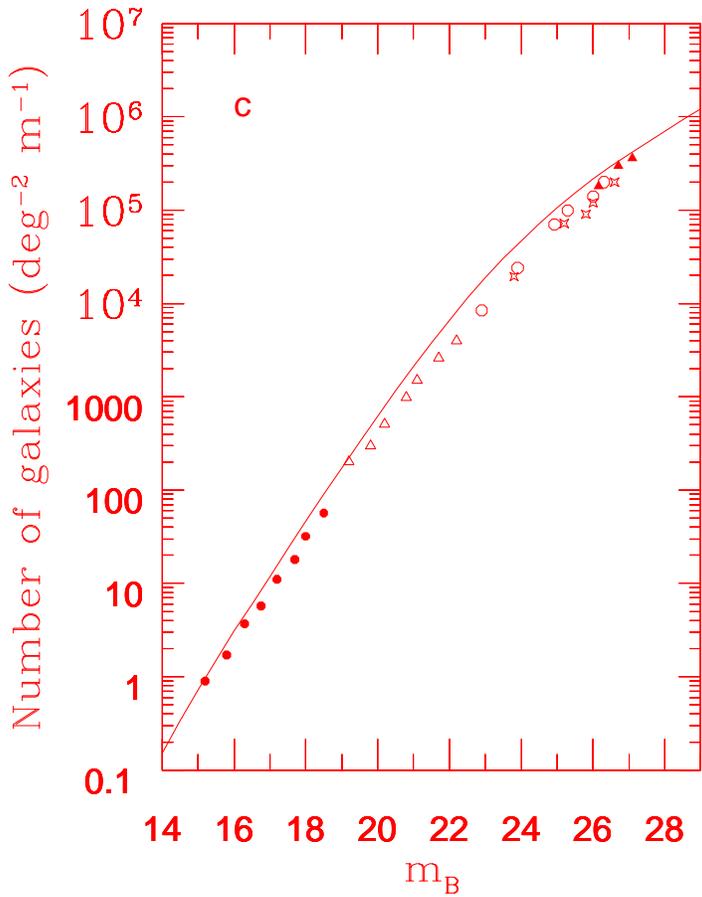
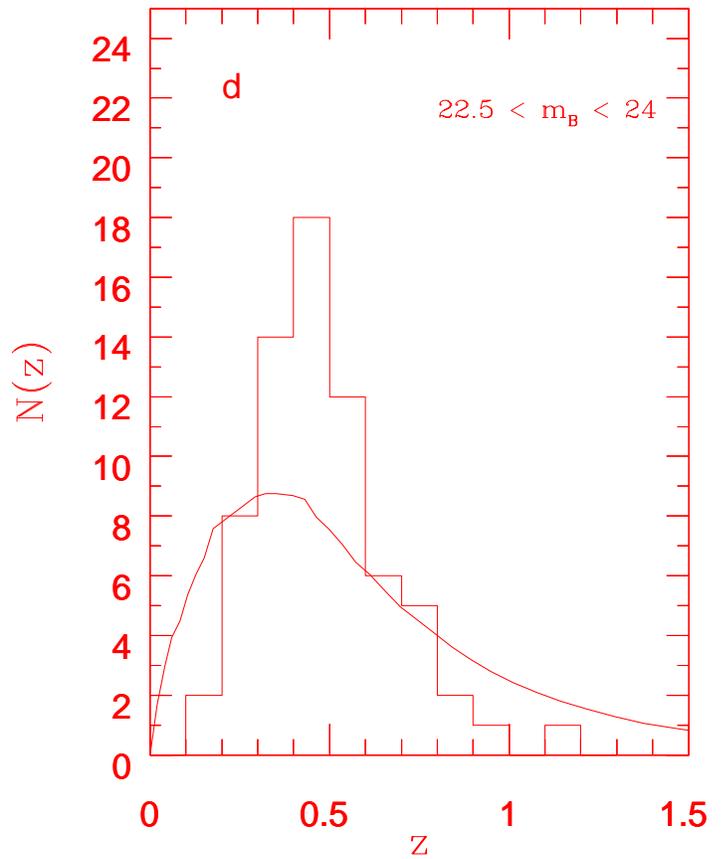
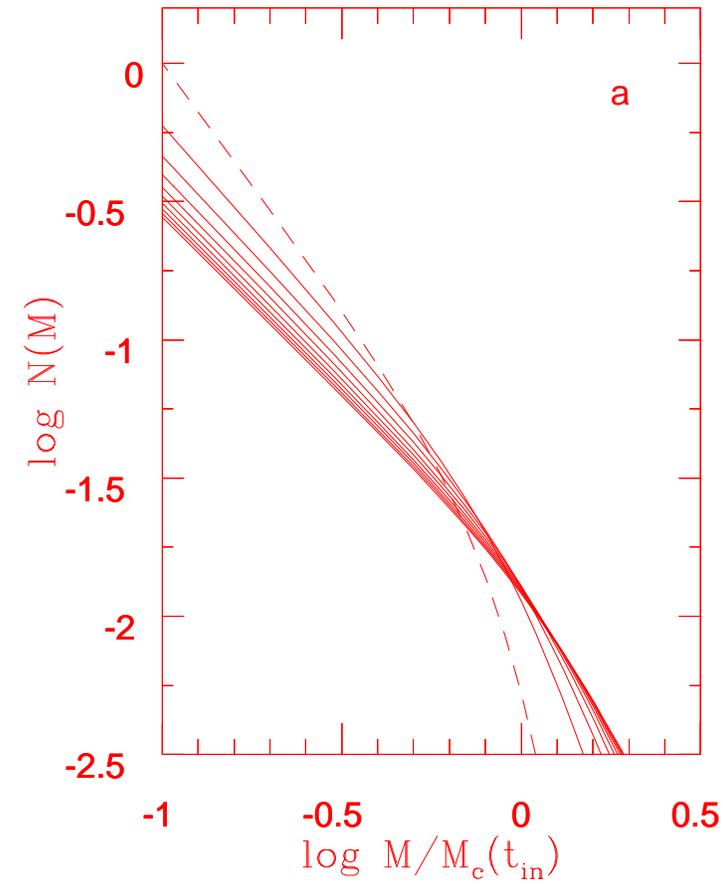
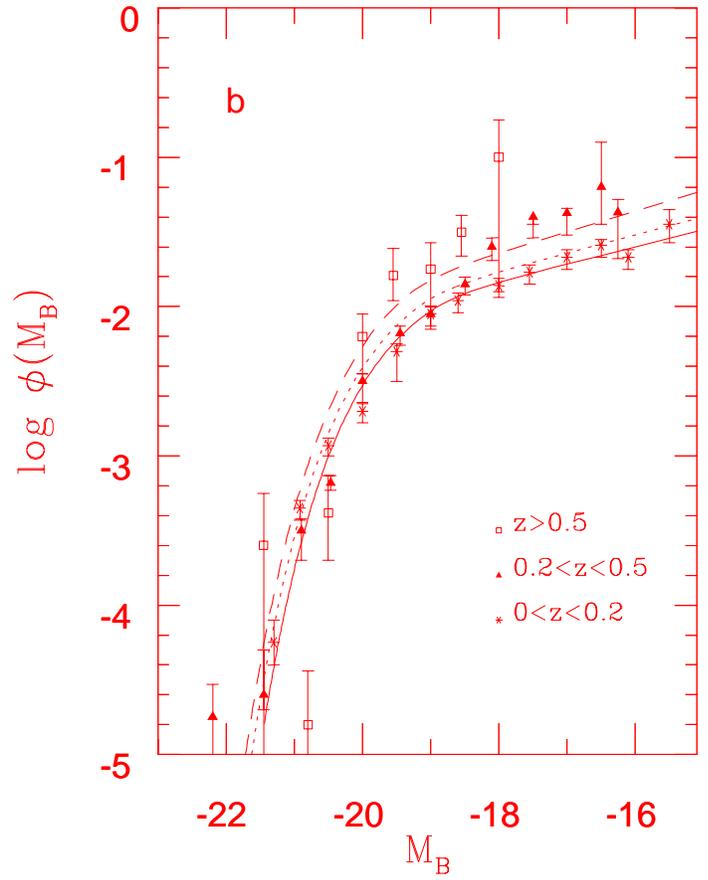
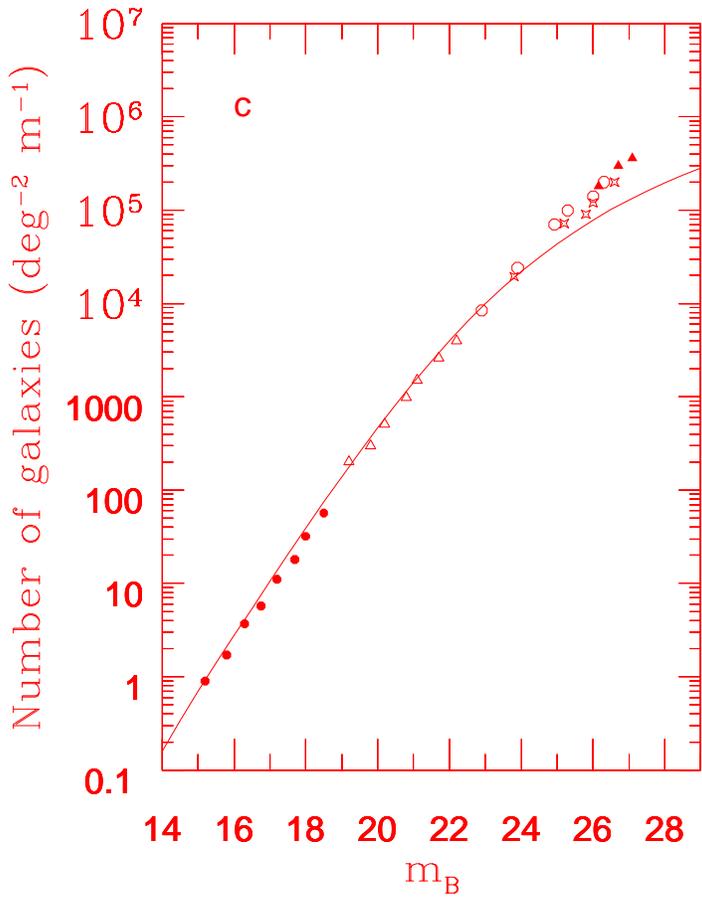
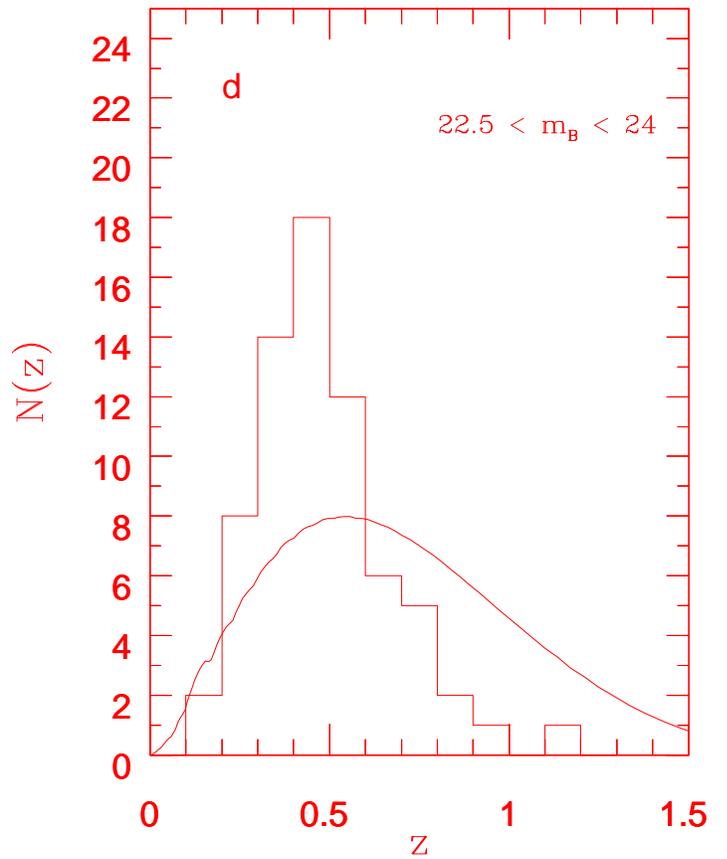
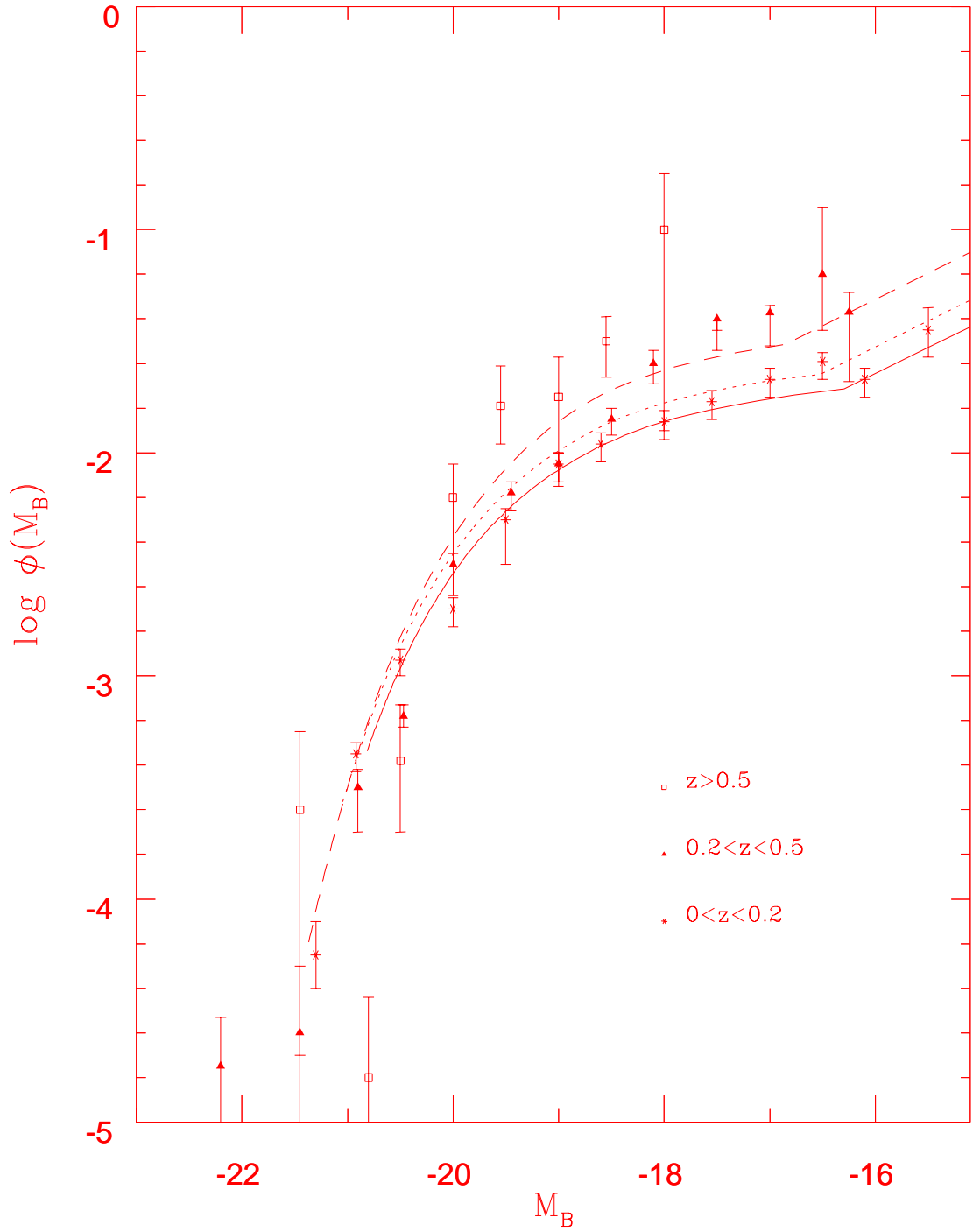
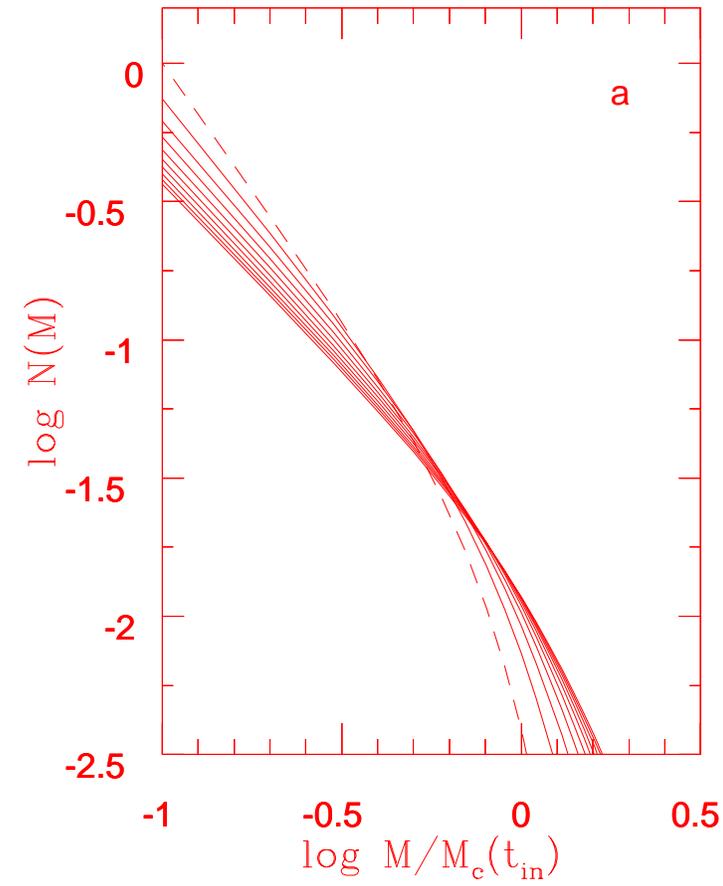
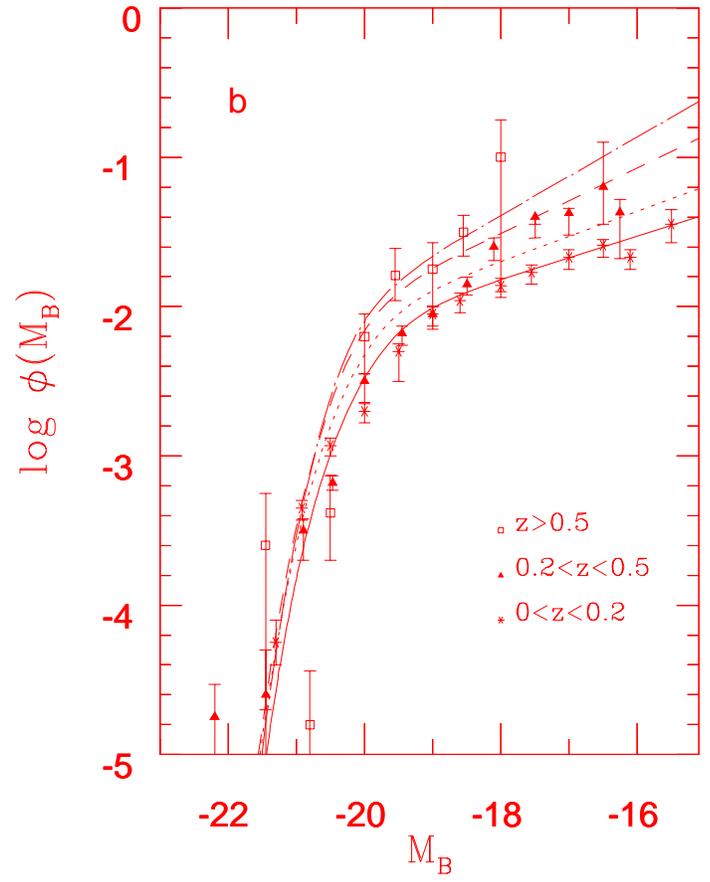
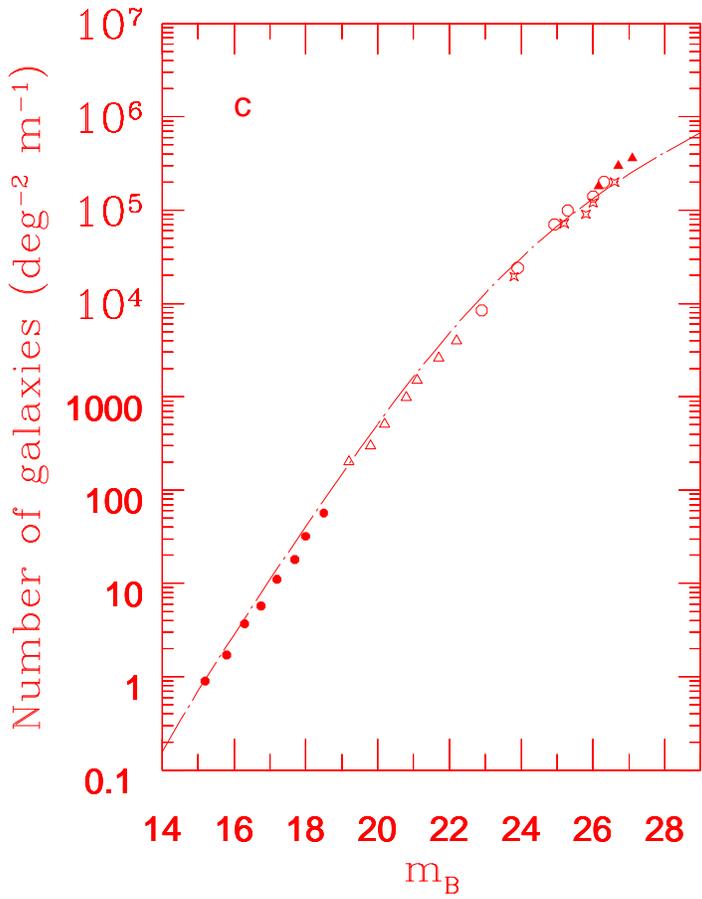
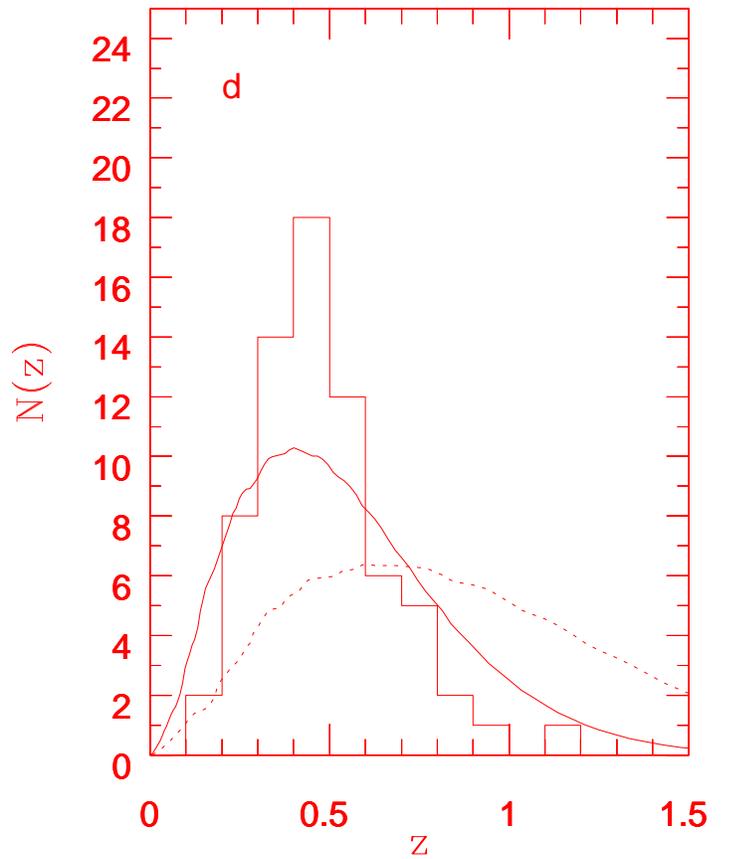
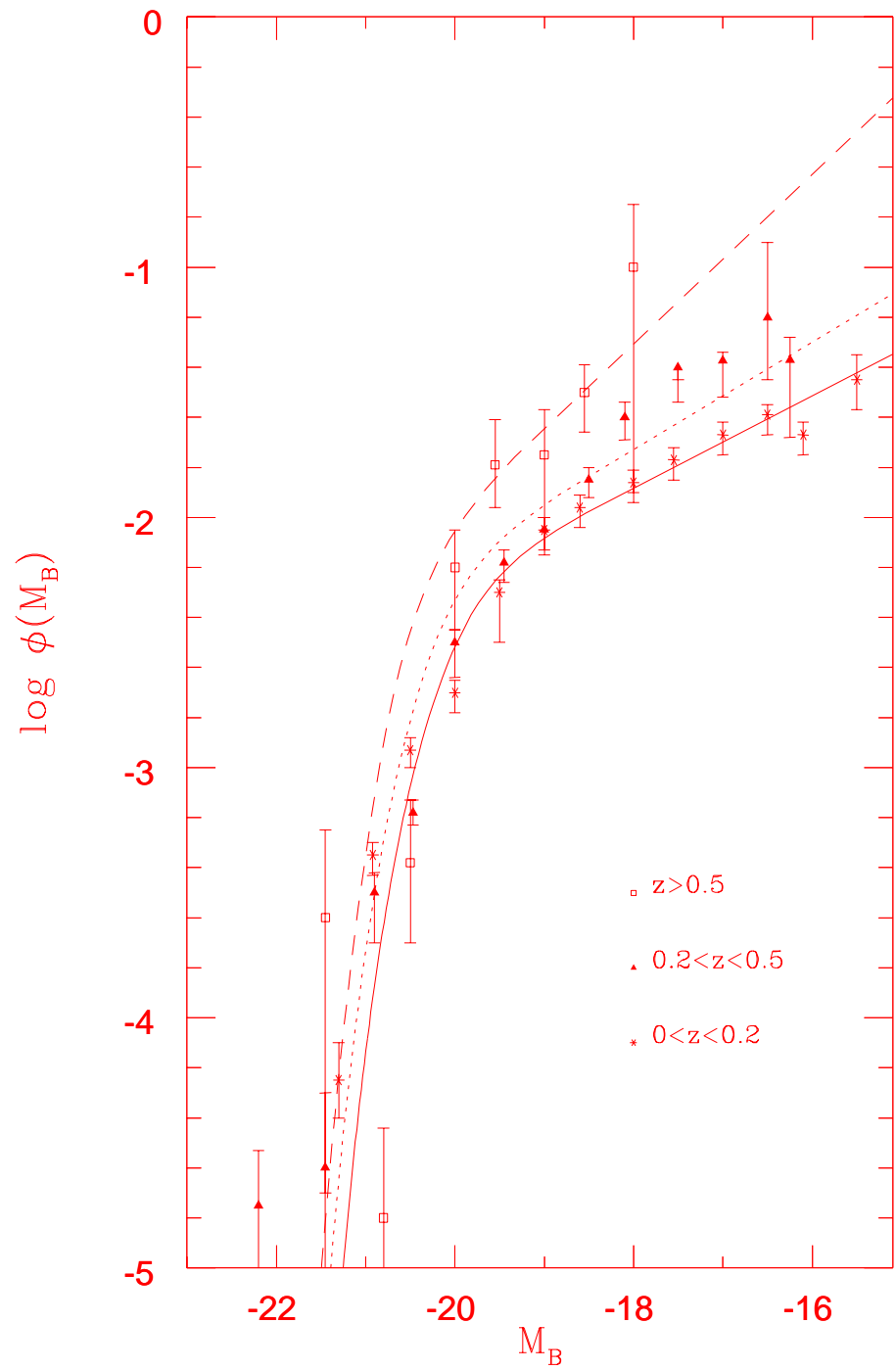
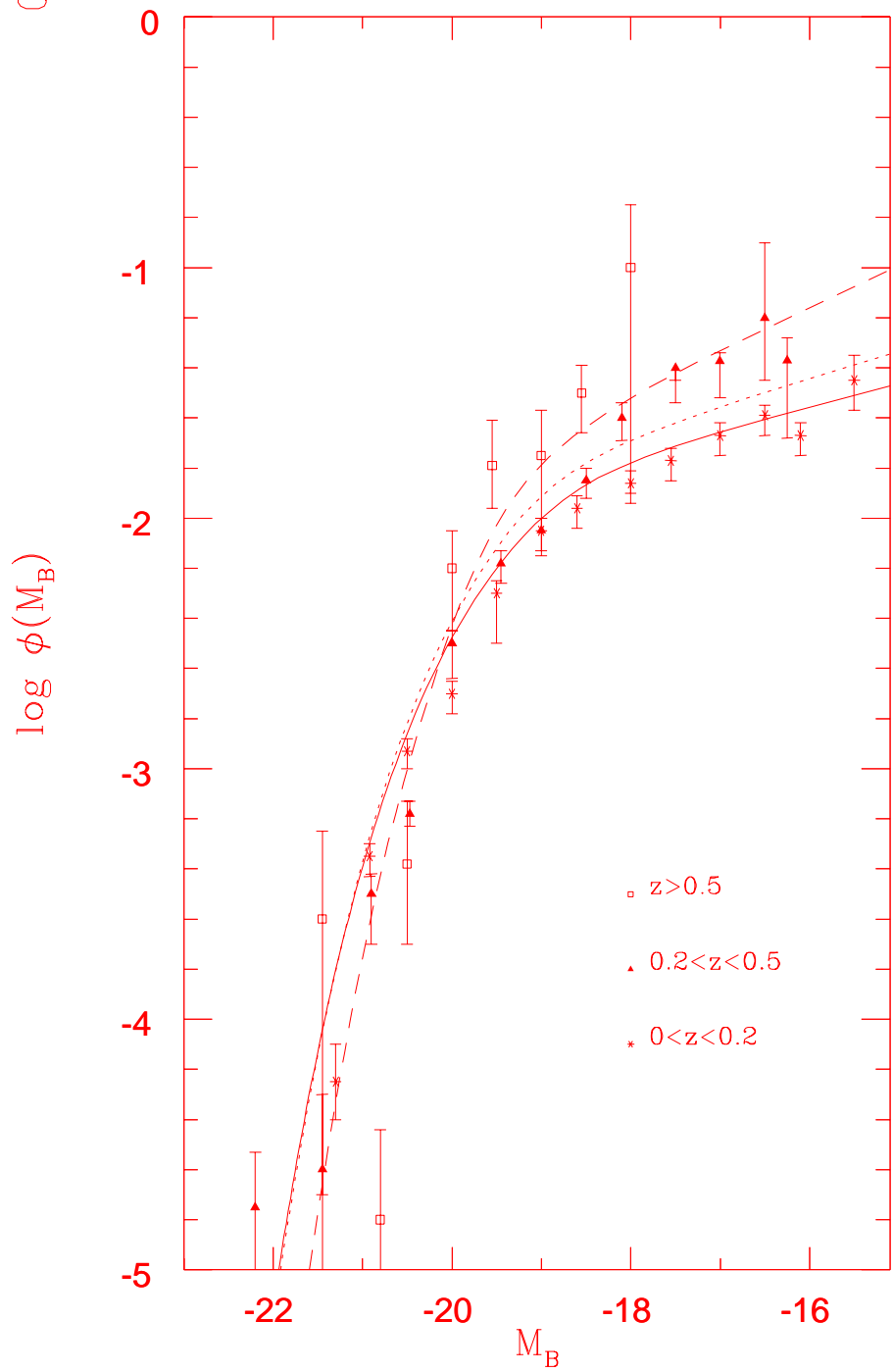
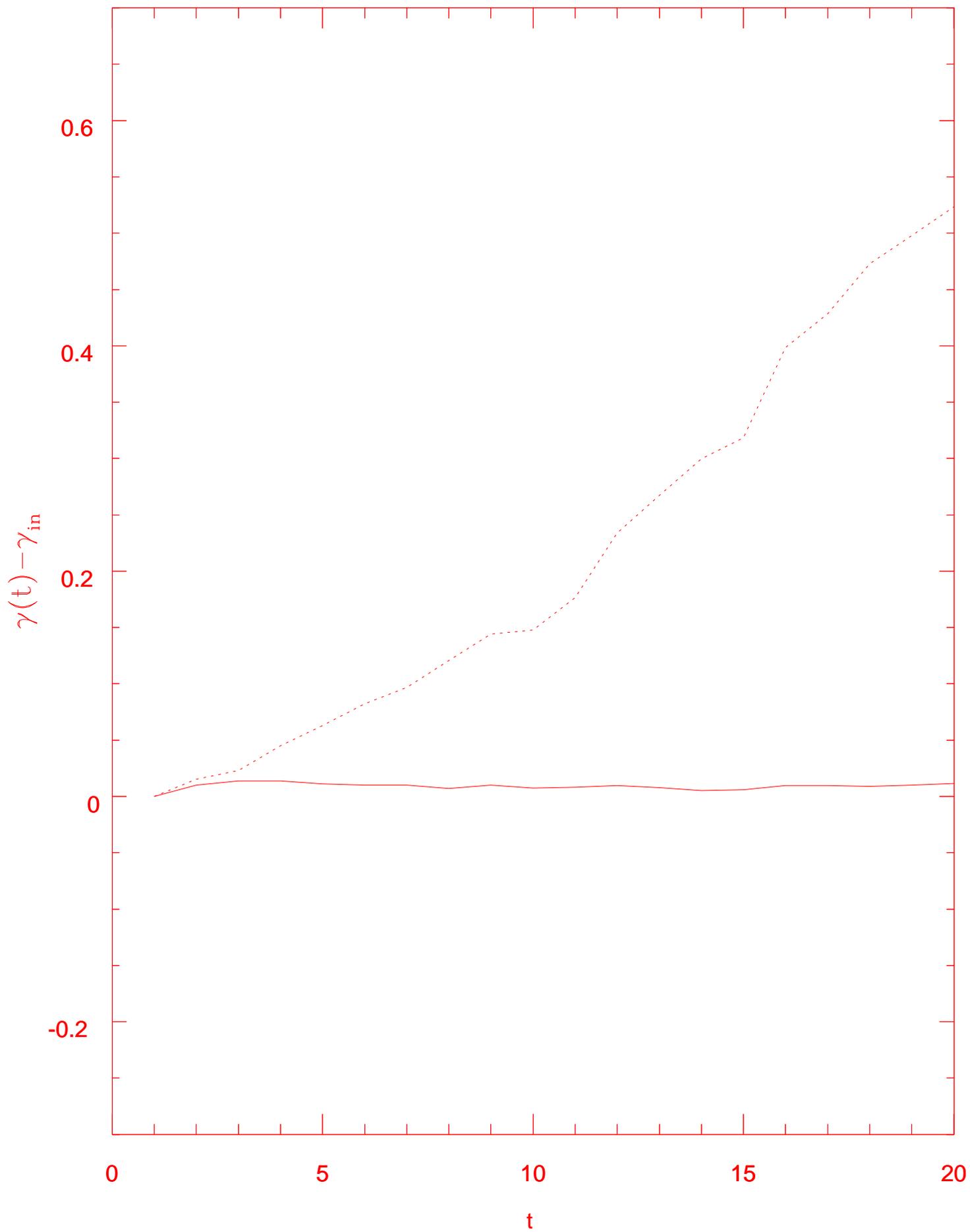
\bye